\documentclass[aps,amsmath,amssymb,superscriptaddress,reprint,10pt,longbibliography, journal=prx]{revtex4-1}

\usepackage{amsmath,amsfonts,amssymb,amsthm,bbm,graphicx,enumerate,times}
\usepackage{mathtools}
\usepackage{tcolorbox}
\usepackage[table,dvipsnames]{xcolor}
\usepackage{colortbl}
\usepackage{comment}
\usepackage{hyperref}
\usepackage{tikz}
\usepackage{graphicx}
\usepackage{textcomp}
\usepackage{siunitx}
\usetikzlibrary{shapes,positioning}
\usepackage{etoolbox}
\usepackage{upgreek}
\usepackage{cleveref}
\usepackage{soul}
\setstcolor{red}

\definecolor{edgeshift}{rgb}{0.881, 0.611, 0.142}
\definecolor{dimple}{rgb}{0.560, 0.692, 0.195}
\definecolor{ramp}{rgb}{0.70, 0.62, 0.85}

  \definecolor{joerg}{rgb}{.0,.5,.0}
  \definecolor{fede}{rgb}{.8,.0,.0}

  \newcommand\tuw{Vienna Center for Quantum Science and Technology (VCQ), Atominstitut, TU Wien, Vienna, Austria}
  \newcommand\ISTA{Institute of Science and Technology Austria, 3400 Klosterneuburg, Austria}

\usepackage{bibunits}
\defaultbibliography{references_arxiv}
\defaultbibliographystyle{apsrev4-1}

\begin{document}

\title{Damping of phonons in one-dimensional quantum fluids}

\author{Federica Cataldini}\email{federica.cataldini@gmail.com}\affiliation{\tuw}%
\author{Nataliia Bazhan}\affiliation{\tuw}
\author{Jo\~{a}o Sabino}\affiliation{\tuw}
\author{Philipp Sch\"{u}ttelkopf}\affiliation{\tuw}
\author{Mohammadamin Tajik}\thanks{Present address: Division of Biology and Biological Engineering, California Institute of Technology, Pasadena, CA, USA}\affiliation{\tuw}%
\author{Frederik S. M{\o}ller}\affiliation{\tuw}\affiliation{\ISTA}
\author{Si-Cong Ji}\affiliation{\tuw}      
\author{Sebastian Erne}\affiliation{\tuw}
\author{Igor Mazets}\affiliation{\tuw} 
\author{J\"{o}rg Schmiedmayer}\email{schmiedmayer@atomchip.org}\affiliation{\tuw}  

\date{\today}

\begin{abstract}
The physics of bosons with repulsive contact interactions in one dimension (1D) admits a range of effective descriptions. In the simplest, the Luttinger liquid (LL) framework, the dynamics reduce to freely propagating, non-interacting elementary excitations (phonons). Beyond this approximation, phonons do interact and therefore experience damping. 
While global relaxation effects have been observed, direct measurements of damping at the level of individual phonons were lacking, leaving the microscopic origin of phonon relaxation unresolved.
In our experiment, we directly and selectively excite individual low-energy phonon modes in a weakly interacting 1D Bose gas and track their time evolution. We observe a non-analytic scaling $\Gamma_k \propto |k|^{\beta}$, with measured exponent $\beta=1.48\,(0.03)$, in excellent agreement with Andreev's prediction $\beta=1.5$, from a self-consistent hydrodynamical approach. This identifies phonon branching and merging as the dominant relaxation mechanism, providing the first direct experimental confirmation of Andreev's universal prediction.
Increasing the excitation strength, we resolve a crossover to a strongly nonlinear regime characterized by wave breaking, directly linking microscopic phonon damping to emergent nonlinear dynamics. 
Our results provide direct access to phonon-level relaxation mechanisms in a nearly integrable quantum system and establish a new experimental paradigm for probing nonlinear and non-equilibrium dynamics in 1D quantum fluids.

\end{abstract}
\maketitle

\begin{bibunit}
\begin{figure*}
\includegraphics[trim={0mm 0mm 0mm 0mm},clip]{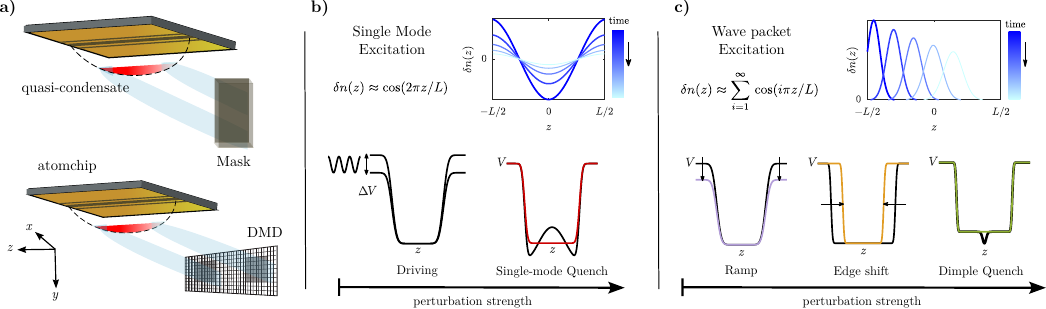}
\caption{\textbf{Schematic of the box trap setup and experimental protocols.} \textbf{(a)} A quasi-condensate is prepared in the harmonic trap generated by the atom chip. A blue-detuned laser beam propagating perpendicular ($x$-direction) to the condensate axis shapes the longitudinal confinement along $z$. Two protocols are used to realize the box potential: either a mask is used to remove the inhomogeneous tails of the harmonic trap (top), or a digital micromirror device (DMD) directly engineers a box-shaped longitudinal potential (bottom).
\textbf{(b)} Protocols used to excite a single phononic mode, illustrated here for the second mode ($i=2$). Excitation is achieved either by resonant driving at frequency $\omega_{dr}=c\pi i/L$, where $c$ is the speed of sound and $L$ is the box length, or by quenching from a potential with an imprinted cosine modulation (single mode potential, in black) to a flat-box potential (red). In both cases, the induced density perturbation takes the form $\delta n(z)\approx\cos(i\pi z/L)$. After excitation, the perturbation undergoes damped oscillatory dynamics (inset).
\textbf{(c)} Protocols used to generate propagating wave packets, corresponding to superpositions of phononic modes. Small-amplitude wave packets are produced by ramping down the initial box potential. Intermediate amplitudes are obtained by adiabatically compressing the box to a shorter length, while large-amplitude wave packets are generated by quenching from a potential containing a central dimple to a flat box. The initial potential is shown in black, whereas the final potentials are shown in purple, orange, and green, respectively. After excitation, the wave packets propagate dispersively through the system (inset). The arrows below panels (b) and (c) indicate the increasing perturbation amplitude associated with each protocol. The box trap realized using the mask is employed for the resonant-driving and ramp protocols, while the DMD-based box trap is used for all other protocols. Additional details are provided in the Methods section.}
\label{fig:schemes}
\end{figure*}

One-dimensional (1D) quantum systems are paradigmatic models of analytically tractable many-body physics, capturing nontrivial interplay between interactions, quantum statistics, and dimensionality~\cite{Girardeau1960,LL1,LL2,Haldane_PhysRevLett.47.1840}. Their theoretical description, from exact diagonalization via bosonization to nonlinear Luttinger-liquid and hydrodynamic descriptions, is well established~\cite{Giamarchi2003,Cazalilla2004,Cazalilla2011,ImambekovRMP2012,GHD_PhysRevX.15.010501}. Ultracold-atom experiments allow direct tests of such theories, enabling quantitative studies of non-equilibrium dynamics and integrability, with key results such as the quantum Newton’s cradle~\cite{Kinoshita2006}, prethermalization~\cite{Gring2012}, long-time recurrences~\cite{Rauer2018}, and tunable dipolar interactions~\cite{LevDDI}, showing how coherence and relaxation emerge in isolated 1D quantum fluids.


The bosonization framework~\cite{Haldane_PhysRevLett.47.1840,Giamarchi2003,Cazalilla2004,Cazalilla2011} captures dynamics in 1D quantum systems in terms of collective bosonic modes.  In the long-wavelength limit ($k \!\to\! 0$), the collective excitations are phonons whose spectrum is linear, $\omega_k = c |k|$, where $k$, $\omega _k$, and $c$ are the wave number of an excitation, its frequency, and the speed of sound, respectively. The resulting Tomonaga–Luttinger liquid theory describes a free fluid, 
in which these phonons are non-interacting and possess infinite lifetimes.  Real systems, however, inevitably deviate from this idealization. Tracing the origins of interactions between collective excitations beyond the Luttinger liquid approximation is one of the central challenges in the field.

Several theoretical approaches have been developed, for example, the nonlinear Luttinger liquid framework deals with band curvature effect, leading to phonon decay and spectral broadening ~\cite{Samokhin1998,Imambekov2009,ImambekovRMP2012}.
A complementary picture is offered by Andreev’s treatment~\cite{Andreev1980}. It derives the phonon's decay rate $\Gamma_k$ from a self-consistent hydrodynamical approach
and predicts a non-analytic $k$-dependence,
$\Gamma_k \propto |k|^{3/2}$. 
Within this framework, phonons acquire a finite lifetime through resonant branching (a phonon with wave number $k$ can decay into two phonons with wave numbers $k_1$ and $k_2=k-k_1$) and merging processes ($k_1$ and $k_2$ merge into a phonon with $k=k_1+k_2$) that satisfy energy and momentum conservation for a continuous set of wave numbers.
Solving this problem in the single-pole approximation for the phonon Green’s function yields an exponential decay of the mode amplitude with a universal non-analytic damping rate
\begin{equation}
\Gamma_k = \alpha \sqrt{\frac{k_B T}{m n_{1\mathrm{d}}}}\, |k|^{3/2},
\label{def:Andreev_damping}
\end{equation}
where $k_B$ is the Boltzmann constant, $T$ is the temperature, $m$ the atomic mass, $n_{1\mathrm{d}}$ the linear density, and $\alpha\sim 1$ is a dimensionless constant, which depends on the details of the mathematical approach; reference~\cite{Andreev1980} yields $\alpha=0.59$. Remarkably, the sound velocity $c$ enters only implicitly through the validity regime $\hbar |k| \!\lesssim\! m c$.  Our work provides the first direct experimental confirmation of this universal prediction by measuring the lifetime of individual phononic modes in a weakly interacting quasi-1D Bose gas of ultracold $^{87}$Rb atoms.\\

Equation~(\ref{def:Andreev_damping}) arises from a hydrodynamic description that represents the classical (mean-field) limit of the Lieb–Liniger model~\cite{LL1}, formulated in terms of conjugate phase and density variables~\cite{MoraCastin}.  In this regime, the interaction of the relevant excitations is captured by a classical field theory or its equivalent hydrodynamical formulation. The underlying mechanism is thus much simpler than in the nonlinear Luttinger-liquid (NLL) framework~\cite{Samokhin1998,Imambekov2009}, where the relevant bosonic quasiparticles emerge as composite pairs of Lieb–Liniger particle- and hole-type excitations with distinct dispersion relations (see Section X of Supplementary Information). It also differs from integrability-breaking mechanisms arising from density-dependent transverse confinement effects~\cite{Mazets2008, Tan2010, Salasnich2002}.  Crucially, the finite damping rate $\Gamma_k \propto |k|^{3/2}$ remains fully compatible with the integrability of the classical mean-field limit~\cite{Zakharov1973}: numerical studies have shown that while phonon populations relax toward stationary distribution, the underlying integrals of motion remain conserved~\cite{Grisins2011}.


The closest experimental evidence to date comes from relaxation of coherently split 1D Bose gases~\cite{Hofferberth2007}, where interference contrast decays as $\propto\exp[-(t/t_0)^{2/3}]$. This was shown theoretically~\cite{Burkov_PhysRevLett.98.200404,Stimming2011} to be the generalized Laplace transform~\cite{Bosch2021} of an ensemble of Andreev-damped modes, consistent with Eq.~(\ref{def:Andreev_damping}). However, this constitutes only indirect evidence: the measurement integrates over all thermally populated modes rather than resolving individual phonon lifetimes, and probes the phase sector rather than the density channel. 
On the other hand, Bragg spectroscopy has previously been used to probe the excitation spectrum of a Bose–Einstein condensate~\cite{steinhauer2002excitation}. In those measurements, however, individual modes were not spectrally resolved and the observed linewidths were dominated by Fourier and inhomogeneous broadening rather than intrinsic damping. Consequently, no information on phonon lifetimes could be extracted.
Direct, mode-resolved access to phonon lifetimes in the density channel has therefore remained the missing experiment, and forms the central advance of the present work.

\begin{figure*}
\includegraphics[trim={0mm 3mm 0mm 1mm},clip]{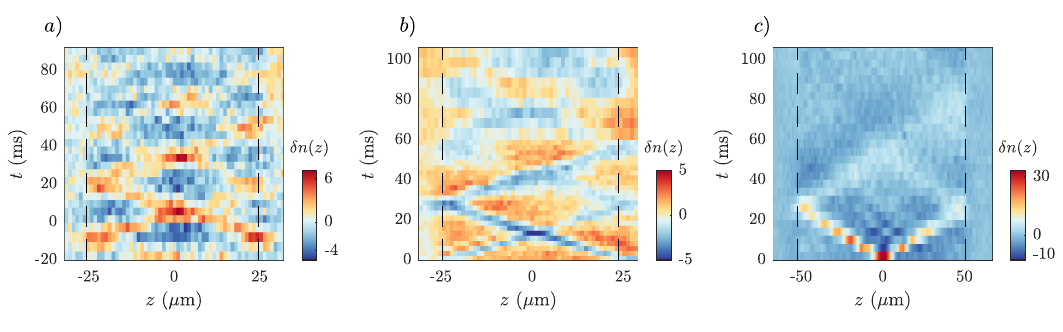}
\caption{\textbf{Density perturbation dynamics.} Time evolution of three distinct density perturbations, referred in the main text as \textit{perturbation carpets}: second mode excited via shaking of the box walls (a), density wave packets emerging from ramping down the potential amplitude in $1$ ms (b), density wave packets emerging after quenching a potential with central dimple to a flat box (c). The x-axis corresponds to the longitudinal direction $z$, while the y-axis shows the evolution time. For panel (a) negative times indicates the driving time, while $t>0$ always represent free dynamics. The position of the box walls during the free evolution is shown with dashed lines. The colorbar indicates the amplitude of the perturbation $\delta n(z)$ in $\upmu$m$^{-1}$. The initial temperature for the three datasets is $64\,(11)\,$, $40\,(4)$ and $69\,(5)$ nK, respectively, while the linear density is $73\,(8)$, $78\,(8)$ and $68\,(7)$ $\upmu$m$^{-1}$.}
\label{fig:Density_carpets}
\end{figure*}

Here, 
we resonantly excite individual phononic modes in a quasi-1D Bose gas and directly measure their coherent dynamics.
This enables excitations with perturbation amplitudes as low as $4\%$ of the linear density, allowing the first direct quantitative test of Andreev's universal hydrodynamic prediction for phonon damping (Eq.~(\ref{def:Andreev_damping})). 

We complement the single-mode measurements with three wave packet excitation protocols that collectively populate low-momentum phononic modes over a controlled range of amplitudes
(Fig.~\ref{fig:schemes}(c)), revealing a crossover from from a regime dominated by phonon branching and merging to a highly nonlinear regime, characterized by wave-breaking.

\begin{figure*}
\centering
\includegraphics[trim={0mm 0mm 0mm 0mm},clip, width=\linewidth]{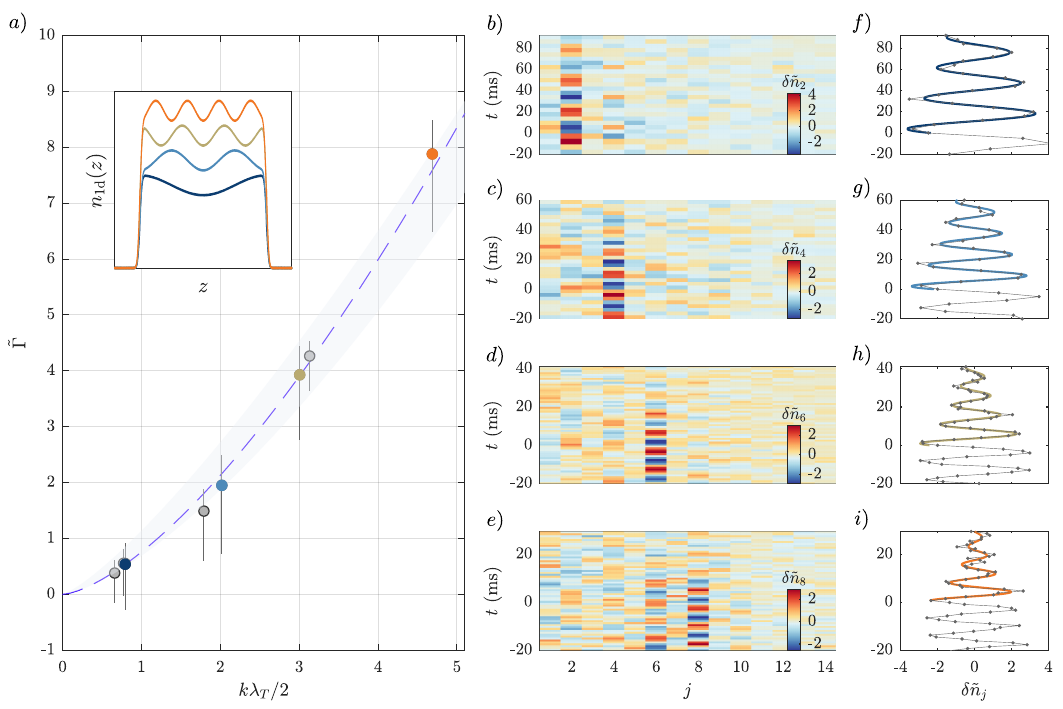}
\caption{\textbf{Damping extracted from single mode dynamics.} \textbf{(a)}: Behaviour of the normalized damping rate $\tilde{\Gamma} = (m\lambda_T^2\,\Gamma)/(4\hbar)$ at different $k$. Experimental data are shown with dots, with errorbars indicating the $68\%$ confidence interval, computed via bootstrap. The dashed line is the curve obtained with fitted coefficient $\alpha = 0.76\,(0.02)$ and fitted exponent $\beta = 1.48\,(0.03)$, where the errors correspond to two standard deviations (see Section VII of Supplementary Information). The shaded area represents the non-polynomial Schr\"{o}dinger equation (NPSE)~\cite{Salasnich2002} simulation of experimental data, taking into account uncertainty on $\tilde{\Gamma}$ due to temperature variation and perturbation amplitude. Four experimental data sets with distinct $k_j$ are highlighted in different colors, and their schematic density profile is shown in the inset of panel (a). Mode $j=2,4,6,8$ are indicated with dark blue, blue, olive green and orange, respectively. \textbf{(b-e)}: Mode decomposition of the selected data sets. It is obtained via Fourier transform of the perturbation profile $\delta n(z,t)$, computed for each evolution time $t$ within the central region of the box (between the vertical dashed lines in Fig. 1) to suppress boundary effects.
On the x-axis is the mode number $j$ and on y-axis is time. Negative values indicate the driving time $\tau_{dr}$; at $t=0$ the modulation is turned off, followed by the free evolution in the box for $t>0$. \textbf{(f-i)}: Full evolution of the addressed mode at a fixed $k$: $k_2 = 2\pi/L$, $k_4 = 4\pi/L$, $k_6 = 6\pi/L$, $k_8 = 8\pi/L$, respectively. The driving time for the four data sets is $\tau_{dr} = 60, 45, 60$ and $40$\,ms, respectively. Of the full driving time only the last $20$\,ms are showed in the plots. Density carpets for modes $4^{th}$ to $8^{th}$ are shown in~\ref{fig_SM:shaking_high_order_density_carpets}.}
\label{fig:main_damping_shaking}
\end{figure*}

In our setup, the quasi-condensate is prepared on an AtomChip~\cite{atomchips}, where a cigar-shaped magnetic trap is overlaid with an optical dipole potential that defines the geometry of the system in the longitudinal direction. 
The experiments are carried out in a box-shaped trap that discretizes the allowed wave number as $k = \frac{\pi}{L}j$, where $L$ is the box size and $j=1,2,3,...$ labels the mode number. 
Note that the discretization of the levels in a box trap does not change the form of the r.h.s. of Eq.(\ref{def:Andreev_damping}), originally obtained~\cite{Andreev1980} in the thermodynamic limit (i.e., for a continuous spectrum of $k$). 
The box trap  in our experiment is realized by shining a blue-detuned laser beam onto an opaque mask, aligned parallel to the longitudinal axis of the atomic cloud, thereby creating two hard walls on the underlying shallow harmonic confinement (see Fig.~\ref{fig:schemes}(a) and Methods).

To resonantly excite the $j^{\mathrm{th}}$ phononic mode (Fig.~\ref{fig:schemes}(b)), we continuously modulate the intensity of the dipole beam at frequency $\omega_{_j}$ and monitor the system’s response through the density perturbation profile recorded at discrete times.
A similar technique was previously used to investigate sound diffusion in a two-dimensional Fermi gas~\cite{Zwirlein_Science2020}.
The density perturbation can be expressed as the sum of density waves $\delta n(z,t) = \sum_{i=1}^\infty\delta\tilde{n}_i(t)\cos(k_iz)$, with $\delta\tilde{n}_i(t)$ the time-dependent amplitude of the $i^{\mathrm{th}}$ mode. Under resonant driving, only the targeted mode contributes significantly, $\delta n(z,t) \approx \delta \tilde{n}_j(t)\cos(k_jz)$.

A representative experiment is shown in Fig.\ref{fig:Density_carpets}(a), where the $2^{\mathrm{nd}}$ mode is excited, yielding $\delta n(z,t) = \delta \tilde{n}_2(t) \cos(2\pi z/L)$. Given the symmetry of the external trapping potential and of the driving protocol, only even modes can be excited; hence the $2^{\mathrm{nd}}$ mode is the lowest we can access. In Fig.\ref{fig:Density_carpets}(a), we show the last $20\,$ms of driving and the subsequent evolution for $100\,$ms. At each time step, we record a statistically large set of absorption images after a $2$ ms time of flight (ToF) expansion, from which we extract the averaged longitudinal density profile $n(z,t)$. The coherent density perturbation is defined as $\delta n(z,t) = n(z,t) - \langle n(z,t)\rangle_t$, where $\langle\cdot\rangle_t$ denotes the time average, which for long enough time, yields the background density.

The single mode measurements presented are performed in a box of length $L\simeq49\,\upmu$m, linear density $n_{1\mathrm{d}}\in[60,75]\,\upmu$m$^{-1}$ and initial temperature $T\in[50,65]$ nK, deep in the 1D regime~\cite{Krueger2010_1Dlimit,Cataldini2022PRX}. The relative perturbation amplitude is $\delta n/n_{1\mathrm{d}} \in [0.04,0.07]$. 

The damping rate of each excited mode is extracted via Fourier analysis of the density perturbation carpet (Fig.~\ref{fig:Density_carpets}(a)), with the resulting mode decomposition shown in Fig.~\ref{fig:main_damping_shaking}(b).
We find that the excited mode is significantly populated and its amplitude $\delta\tilde{n}_2(t)$ evolves following a damped oscillation (Fig.~\ref{fig:main_damping_shaking}(f)), whereas the amplitude of higher modes is negligible, demonstrating that we can excite a single phononic mode with great accuracy. 

In addition to the $2^{\mathrm{nd}}$ mode we are able to excite the next three symmetric modes (see Fig.~\ref{fig_SM:shaking_high_order_density_carpets}), thus spanning a large interval of $k$-space. The mode decomposition for measurements where the $4^{\mathrm{th}}$, $6^{\mathrm{th}}$, and $8^{\mathrm{th}}$ modes are excited is shown in Fig.\ref{fig:main_damping_shaking}(c-e). We observe that the phonons' lifetime $\tau = \Gamma^{-1}$ decreases for higher-order modes, and it can be extracted by fitting $\delta\tilde{n}_j(t)$ with the function: 
\begin{equation}
f(t) = A\exp(-t/\tau)\cos(\omega t\,+\,\phi),
\label{eq:fit_function}
\end{equation}
where $A$, $\omega$ and $\phi$ are the other fitting parameters, representing the initial amplitude of the mode, oscillation frequency, and initial phase, respectively. 

To explore the $k$-dependence of the damping, we analyze nine independent measurements, each addressing one of the four lowest modes, and extract the corresponding damping rate. Since the measurements were performed under slightly different experimental conditions, specifically variations in linear density and initial temperature, we compare them in Fig.~\ref{fig:main_damping_shaking}(a) using normalized quantities, allowing for a direct and consistent comparison across data sets: We use $\tilde{\Gamma} = \frac{m\lambda_T^2}{4\hbar}\Gamma$ and $\tilde{k} = k\lambda_T/2$, with $m$ being the mass of $^{87}$Rb atom and $\lambda_T = 2\hbar^2n_{1\mathrm{d}}/(mk_BT)$ the thermal coherence length.

Fitting the extracted damping with a power law yields an exponent $\beta = 1.48\,(0.03)$, in excellent quantitative agreement with Andreev's prediction (Eq.~\ref{def:Andreev_damping}). The prefactor $\alpha = 0.76\,(0.02)$ lies slightly above the mean-field value $\alpha = 0.59$. The latter is not expected to be quantitatively accurate, due to the single-pole ansatz and the vertex truncation in~\cite{Andreev1980}; the $\sim25\%$ agreement is therefore better than anticipated.
 

The non-analytic behavior $\Gamma_k \propto |k|^{3/2}$ as $k \rightarrow 0$ establishes phonon merging and branching as the dominant relaxation channel in weakly interacting one-dimensional Bose gases. Importantly, this result is in clear contrast to the treatment applicable to higher dimensions \cite{Hohenberg1965,Szepfalusy1974, PITAEVSKII1997398}, deeper in the degenerate regime than in Refs.~\cite{Cornell_PhysRevLett.78.764, Dalibard_PhysRevLett.121.145301}.

\begin{figure*}
\centering
\includegraphics[trim={0mm 0mm 0mm 2mm},clip]{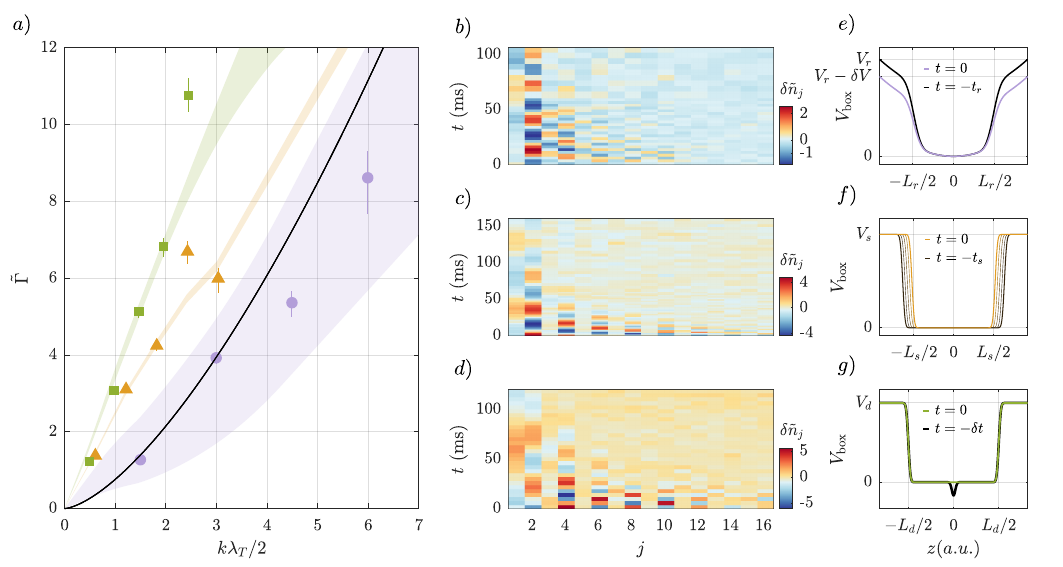}
\caption{\textbf{Damping extracted from wave packets dynamics.} Panel a): behaviour of $\tilde{\Gamma}$ for different $\tilde k=k\lambda_T/2$. Data points of the same color represent the lower modes excited through different protocols: fast ramp (purple circles), symmetric displacement of box walls (orange triangles), quench of a potential with central dimple to a flat box (green squares). Errorbars indicate the $68\%$ confidence interval, computed via bootstrap. The relative initial temperature is $T_r = 40\,(4)$ nK, $T_s = 60\,(6)$ nK and $T_d = 69\,(5)$ nK, where $r$, $s$, and $d$ label the ramp, edge-shift, and dimple experiments, respectively.
The mode decomposition for the purple, orange and green data set are displayed in panel b), c) and d), respectively, with colormap giving the amplitude of the $j^{\mathrm{th}}$ mode, $\delta\tilde n_j$. The corresponding evolution of the density perturbation in real space is show in Fig.\ref{fig:Density_carpets}b), Fig.~\ref{fig:wp_wave_breaking}a) and Fig.\ref{fig:Density_carpets}c). The black solid line is the expected Andreev damping obtained with coefficient $\alpha = 0.76$ as obtained from the fit of the shaking measurements (Fig.~\ref{fig:main_damping_shaking}a)). The shaded areas represent finite temperature NPSE simulations of experimental data, taking into account uncertainty on temperature and on perturbation amplitude. Panels e) to g): Schematic illustration of the three protocols; relative proportions are not to scale.}
\label{fig:main_damping_wp}
\end{figure*}

The single-mode measurements access the regime of weak excitations, where the finite lifetime of phonons arises from leading-order nonlinear hydrodynamic processes (see Section IX of Supplementary Information). A natural next question is how this picture evolves as the perturbation strength increases and higher-order nonlinearities become significant. To explore this, we move beyond single-mode excitation and consider wave packets, which consist of controlled superpositions of phononic modes.

We implement three different protocols that enable systematic tuning of the perturbation amplitude, ranging from $5\%$ up to $40\%$ of the background density. This approach allows us to experimentally investigate the crossover from a regime dominated by phonon damping to one in which higher-order nonlinear effects, such as wave steepening and breaking, govern the dynamics, while remaining within the same hydrodynamic framework.

The initiation of wave packets excites several low-lying modes with frequencies $\omega_j = c k_j = c\pi j/L$, with the finite imaging resolution ($\sim1.5\,\upmu$m) limiting the observable mode number to $j_{max}\approx 12-18$, depending on the system size.

The low amplitude regime is investigated by ramping down the box potential by a small quantity $V_{box}(t\ge0) = V_{box}(t=-t_r) - \delta Vr$, where $t_r=1$ ms is the short ramping time, $V_{box}(t=-t_r)\equiv V_r$ is the potential at the thermal state and $\delta V$ is the small potential variation (Fig.~\ref{fig:main_damping_wp}(e)). Here and in the following protocols, we keep the convention that $t=0$ denotes the onset of the free evolution.
An experiment where the initial potential is ramped down is shown in Fig.~\ref{fig:Density_carpets}(b), where a perturbation amplitude of about $5\%$ of the linear density $n_{1\mathrm{d}} = 78\,(8)$ $\upmu$m$^{-1}$ is achieved. Its mode decomposition is plotted in Fig.\ref{fig:main_damping_wp}(b). This protocol is well suited for accessing the weak-perturbation regime, as it enables controlled small-amplitude density modulations. However, its applicability is limited when targeting larger perturbations: increasing the modulation strength leads to deviations from linear control of the trapping potential. This constrains achievable amplitudes and motivates the use of alternative protocols to access stronger perturbation strengths.

\begin{figure}
\centering
\includegraphics[trim={0mm 2mm 0mm 0mm},clip]{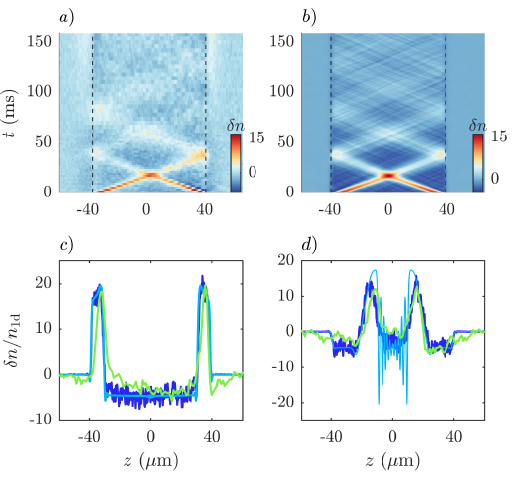}
\caption{\textbf{Comparison between NPSE simulations and experimental data of a large amplitude wave packet.} Panel a): experimental data for the evolution of a density perturbation which is about $20\%$ of the linear density $n_{1\mathrm{d}} = 78\,(8)\,\upmu$m$^{-1}$. Panel b): NPSE simulations at $T=65$ nK and $n_{1\mathrm{d}} = 75\,\upmu$m$^{-1}$. Vertical dashed lines represent the position of the box walls. Panel c) and d): density perturbation profiles $\delta n/n_{1\mathrm{d}}$ at $t=0$ (left) and $t=10$ ms (right). Experimental profile is plotted in green, in dark (light) blue is the profile obtained via finite (zero) temperature NPSE simulations.}
\label{fig:wp_wave_breaking}
\end{figure}

At intermediate amplitudes, wave packets are generated by slightly shrinking the box (Fig.~\ref{fig:main_damping_wp}(f)), shifting each wall by $\Delta L \sim 2\,\upmu$m over $t_s = 5$ ms: $L(t>0) = L(t=-t_s) -2\Delta L$. The mechanism induces two counter-propagating wave packets, whose density perturbation is $\sim20\%$ of the linear density $n_{1\mathrm{d}} = 78\,(8)$ $\upmu$m$^{-1}$. This protocol is used for the experiment shown in Fig.~\ref{fig:wp_wave_breaking}(a), with the corresponding mode dynamics displayed in Fig.~\ref{fig:main_damping_wp}(c).

To achieve large perturbation amplitudes, up to $40\%$, we employ a third protocol based on a quench from a box potential with a central dimple to a flat box (Fig.~\ref{fig:main_damping_wp}(g)). In the previous approach, increasing the edge displacement leads to strong excitation of higher-momentum components and rapid dispersion, preventing access to this regime. In contrast, the dimple quench generates two wave packets that propagate symmetrically from the center toward the edges with opposite velocities, while maintaining a well-defined structure. By varying the initial dimple size, we achieve precise control over the perturbation amplitude.
A typical experiment is shown in Fig.~\ref{fig:Density_carpets}(c): Each wave packet has an amplitude corresponding to $39\%$ of the linear density $n_{1\mathrm{d}} = 68\,(7)$ $\upmu$m$^{-1}$ (Fig.~\ref{fig_SM:wp_dimple_scheme}). 

The latter two protocols (Fig.~\ref{fig:main_damping_wp}(f-g)) are implemented by replacing the opaque mask with a digital micromirror device (DMD) which enables arbitrary shaping of the longitudinal potential (see Fig.~\ref{fig:schemes}(a) and Methods). 
Using the DMD offers the advantage of compensating for the residual shallow harmonic potential generated by the chip \cite{Tajik:19}, where the quasi-condensate is prepared, thereby improving the symmetry and homogeneity of the system.
We notice indeed that, the population of odd modes in the two experiment realized with DMD is overall smaller than in the low-amplitude experiment. 
In general, non-resonant mode populations arise primarily from thermal fluctuations, with small contributions from potential asymmetries. Finite phonon lifetimes broaden the resonance lines, $\Delta \omega \sim \Gamma_k$, explaining the weak excitation of non-resonant modes (such as the sixth harmonic during eighth mode driving) consistent with nonlinear mode mixing. 

The extracted damping for the ensemble of modes excited through wave packet initiation is shown in Fig.~\ref{fig:main_damping_wp}(a). Here we plot the normalized damping $\tilde{\Gamma}$ as a function of $\tilde{k}$, for each of the three experiments: low-amplitude perturbations (purple), and stronger perturbations induced by edge displacement (orange) and dimple quench (green). 
Due to the finite system size and the imaging resolution, only a limited number of modes is accessible, in the given range of linear density ($70-90\,\upmu$m$^{-1}$). Therefore, the collected data do not form a Gaussian statistical distribution, thus preventing a meaningful fit. Instead, we directly compare the measurements with the theoretical damping expression of Eq.(~\ref{def:Andreev_damping}) (black line), where we have chosen $\alpha=0.76$, as obtained from the single mode analysis (Fig.~\ref{fig:main_damping_shaking}(a)). 
Interestingly, the observed damping at weak perturbation reproduces the behaviour predicted by Andreev's hydrodynamic theory, with the previously found coefficient, while it significantly deviates from it for stronger perturbations. A more detailed analysis to ensure statistical significance to our measurements is provided in Section VIII of Supplementary Information. 


To investigate the dynamics of the 1D quasi-condensate in the regime of large perturbation amplitude, we provide, in Fig.~\ref{fig:wp_wave_breaking}(a) a more detailed comparison between experiment and numerical simulations. We consider the dynamics of the wave packet obtained via edge displacement, whose perturbation amplitude is about $20\%$ of $n_{1\mathrm{d}}$, and we find that the observed dynamics are in excellent agreement with the simulated data (Fig.~\ref{fig:wp_wave_breaking}(a,b)). 

To gain more insight about the underlying physics, we look at the profile of the density perturbation at two distinct times, $t=0$ which sets the end of the walls displacement and $t=10$ ms. In both cases, we find a perfect overlap between the experimental profile (green line) and the NPSE data (dark blue line). In addition to finite temperature NPSE numerical simulation, we consider NPSE simulations at zero temperature (light blue line) and compare them to experimental data: while the agreement at $t=0$ is still good, after $10$ ms we notice that the zero-temperature profile features fast oscillations, suggesting that a wave-breaking process has occurred \cite{shockwave, schuttelkopf2024}. These high-frequency oscillations are of the order of the healing length, hence below our imaging resolution.
The phenomenon is a consequence of the velocity gradient $\Delta c = c_p - c_b$ between the base ($b$) and the peak ($p$) of the propagating wave packet.  This leads to a deformation of the wave packet, with the peak moving faster than the base. 
Whenever $\Delta c\,t$ becomes on the order of the characteristic size of the wave packet ($\sigma_{wp}$) a shock wave forms. For $\frac{\Delta n}{n_{1\mathrm{d}}} \ll 1$, $\Delta c = \frac{dc}{dn}\Delta n = \frac{c}{2}\frac{\Delta n}{n_{1\mathrm{d}}}$ and the typical timescale for wave breaking $t_{wb}$ can be estimated as
\begin{equation}
    t_{wb}^{-1} \approx \frac{c}{2\sigma_{wp}}\bigg(\frac{\Delta n}{n_{1\mathrm{d}}}\bigg).
    \label{eq:wave_braking_time}
\end{equation}
In the equation, $\Delta n$ is the initial amplitude of the perturbation, $\Delta n=\delta n(t\approx0)$. For the wave packets shown in Fig.~\ref{fig:wp_wave_breaking}(a), with $\Delta n/n_{1\mathrm{d}}\sim 20\%$, $c=1.9\,$ $\upmu$m/ms and $\sigma_{wp}\equiv w_{R}\approx w_L \approx 2 \, \mathrm{\upmu}$m, we estimate $t_{wb}\approx 10\,$ ms, which is shorter than the observed evolution time and broadly consistent with numerics (Fig.~\ref{fig:wp_wave_breaking}(d)). 
The wave-breaking mechanism occurs at both zero and finite temperatures, 
though thermal fluctuations wash out the high-frequency oscillations~\cite{shockwave}; its signature nevertheless remains visible in the coherent dynamics as a deviation from $|k|^{3/2}$ damping, as seen in Fig.~\ref{fig:main_damping_wp}(a). 


\begin{figure}
\centering
\includegraphics[trim={0mm 0mm 0mm 0mm},clip]{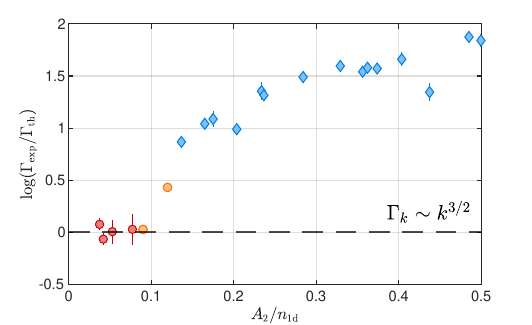}
\caption{\textbf{Crossover from weakly to highly nonlinear regime.}  Measured damping rate of the second mode, expressed as $\Gamma_{\mathrm{exp}}/\Gamma_{\mathrm{th}}$, plotted versus the relative mode amplitude $A_2 / n_{1\mathrm{d}}$, with $A_2$ the mode amplitude and $n_{1\mathrm{d}}$ the linear density. The vertical axis shows $\log\left(\Gamma_{\mathrm{exp}}/\Gamma_{\mathrm{th}}\right)$ to enhance the visibility of the data. The theoretical damping $\Gamma_{\mathrm{th}}$ is given by Eq.~(\ref{def:Andreev_damping}) with $\alpha = 0.76$. Circles (diamonds) denote excitation via shaking (quench of the trapping potential). Red points correspond to the data shown in Fig.~\ref{fig:main_damping_shaking}, while orange points indicate higher-amplitude data excluded from that analysis. Datasets in diamond have linear density $n_{1\mathrm{d}}\in[60,90]\,\mathrm{\upmu m^{-1}}$ and temperature $T\in[50, 130]$ nK (see Table~\ref{T_sm:PB_meas_list} of Supplementary Information). The dashed line marks $\Gamma_{\mathrm{exp}} = \Gamma_{\mathrm{th}}$. Errorbars correspond to the $68\%$ confidence interval, computed via bootstrap.}
\label{fig:gamma2_vs_perturbation_amp}
\end{figure}

Although wave breaking can occur even for weak perturbations, the associated timescale is much longer, allowing Andreev damping to dominate at small amplitudes. Taken together, these results identify phonon branching and merging as the dominant relaxation channel in the weak-excitation regime, and wave steepening and breaking as the governing mechanism at large amplitudes.

A natural question is whether these two mechanisms can be placed on a single quantitative axis. Figure~\ref{fig:gamma2_vs_perturbation_amp} provides this comparison, focusing on the second mode. The relevant observable is the relative perturbation amplitude $A_2/n_{1\mathrm{d}}$, where $A_2$ is the amplitude of the second mode obtained from the fit (Eq.~\ref{eq:fit_function}). By measuring its damping rate across excitation strengths ranging from $4\%$ to $50\%$ of the background density, we track how the system evolves between two physically distinct regimes within a single experimental platform. At weak excitation, the ratio between experimental and theoretical damping, $\Gamma_{\mathrm{exp}}/\Gamma_{\mathrm{th}}$, clusters near unity. Here, $\Gamma_{\mathrm{th}}$ is computed from Eq.~\eqref{def:Andreev_damping} using $\alpha = 0.76$ obtained from the single-mode analysis (Fig.~\ref{fig:main_damping_shaking}(a)).  Above approximately $10\%$ perturbation amplitude, a systematic deviation emerges and grows monotonically, consistent with the onset of wave-breaking dynamics identified in Figs.~\ref{fig:main_damping_wp} and~\ref{fig:wp_wave_breaking} and reproduced by our finite-temperature NPSE simulations.

In the figure, circles denote measurements where the mode is excited via resonant driving. This protocol is not suitable for reaching large perturbations, since stronger driving leads to substantial atom loss as particles are expelled from the trap. To extend to larger amplitudes, we use different independent datasets (diamonds) in which the second mode is populated by imprinting a cosine potential $V_c \propto \cos(2\pi z/L)$ followed by a quench to a flat box, as shown in Fig.~\ref{fig:schemes}(b).

These large-amplitude datasets were previously used as a test ground for Generalized Hydrodynamics (GHD)~\cite{Cataldini2022PRX}, where the long-time relaxation was successfully described by ballistic dephasing of rapidity distributions, a qualitatively distinct mechanism. Figure~\ref{fig:gamma2_vs_perturbation_amp} places these two groups of measurements on a common axis and shows that they form a smoothly connected family of experimental observations. 
How the two theoretical descriptions, exponential damping in the weakly perturbed hydrodynamic limit and ballistic rapidity dephasing in the strongly perturbed regime connect is an open theoretical question. 

In this sense, Fig.~\ref{fig:gamma2_vs_perturbation_amp} serves both as a synthesis and as an outlook of this work. As a synthesis, it places the microscopic damping regime and the strongly nonlinear regime on a single quantitative axis within one experimental platform. As an outlook, it identifies a concrete open theoretical challenge: a unified  framework that describes both limiting behaviors, and the smooth crossover between them. \\

In summary, we have achieved the first direct, mode-resolved observation of phonon damping in a one-dimensional Bose gas. By selectively exciting individual phononic modes, we revealed the characteristic $|k|^{3/2}$ scaling of the damping rate predicted by Andreev's hydrodynamic theory, identifying phonon branching and merging as the dominant relaxation channel in the weak-excitation regime. By extending to wave packet excitations of controlled amplitude, we resolved the crossover to a strongly nonlinear regime governed by wave breaking. The crossover is characterized quantitatively in Fig.~\ref{fig:gamma2_vs_perturbation_amp}.

The most immediate theoretical challenge raised by this work is the relation between Andreev's hydrodynamic damping and the GHD description of the same near-integrable system. The data of Fig.~\ref{fig:gamma2_vs_perturbation_amp} demonstrate that both regimes are accessible in a single experimental platform and connect smoothly as a function of perturbation amplitude.

Several experimental directions follow naturally. Connected higher-order phase correlations~\cite{Schweigler2017Nature,Zache2020PRX} would provide direct access to the three-phonon vertex underlying the branching and merging processes, complementing the lifetime measurements presented here. Systematic variation of temperature at fixed $k$, particularly accessible in optical-lattice realizations of the Lieb-Liniger model, would test the predicted $T^{1/2}$ scaling. Extension toward higher momentum, where $\hbar k$ approaches $mc$, would probe the crossover from hydrodynamic damping to the universal power-law edge singularities predicted by nonlinear Luttinger-liquid theory~\cite{ImambekovRMP2012,Imambekov2009}. Improved control over system stability would allow the investigation of non-mean-field effects beyond the Ehrenfest time at long evolution times~\cite{Mazets_2026}. Together, these directions outline a quantitative experimental program for testing microscopic theories of relaxation in one-dimensional quantum fluids.

%

\end{bibunit}

\newpage
\clearpage
\begin{bibunit}

\newpage
\onecolumngrid
\large{\bf{Methods and Supplementary Information}}\\
\normalsize

\setcounter{figure}{0}
\renewcommand{\thefigure}{SI\arabic{figure}} 
\renewcommand{\theHfigure}{SI.\arabic{figure}}

\section{Phononic regime.}
\begin{figure*}[b]
\centering
\includegraphics[trim={0mm 0mm 0mm 0mm},clip]{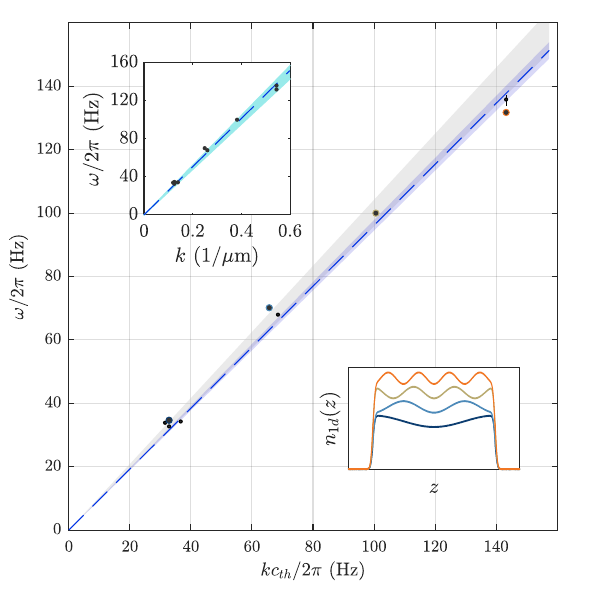}
\caption{\textbf{Dispersion relation from single mode measurements.} The excitation frequency $\omega/2\pi$, extracted from fits to the mode dynamics (Fig.~\ref{fig:main_damping_shaking}(f–i)), is plotted against the expected frequency $\omega_{th} = c_{th}k$, calculated from the measured linear density. Here, $c_{th}=\sqrt{g_n n_{1\mathrm{d}}/m}$ is the effective density-dependent interaction strength, $n_{1\mathrm{d}}$ is the linear density and $m$ is the atomic mass. Highlighted points correspond to the measurements shown in the main text, with colours indicating different mode numbers as defined by the scheme in the lower inset. The dashed blue line shows a linear fit to the data, yielding a speed of sound $c_{fit}=1.59\,(0.03)\,\mathrm{\upmu}$m/ms; the shaded blue region denotes the fit uncertainty. The grey shaded region indicates the expected range of $\omega_{th}$, corresponding to the measured density interval $n_{1\mathrm{d}}\in[60-75]\,\mathrm{\upmu}$m$^{-1}$. The fitted and expected speeds of sound agree within uncertainties. \textit{Upper inset:} Excitation frequency $\omega/2\pi$ as a function of the momentum $k$, with linear fit (dashed blue line) and NPSE simulations (turquoise shaded region). \textit{Lower inset:} Schematic of the lowest four symmetric modes imprinted on the density profile, from the second mode (dark blue) to the eighth (orange).
}
\label{fig_SM:dispersion_shaking}
\end{figure*}

\begin{figure*}
\centering
\includegraphics[trim={10mm 0mm 0mm 0mm},clip]{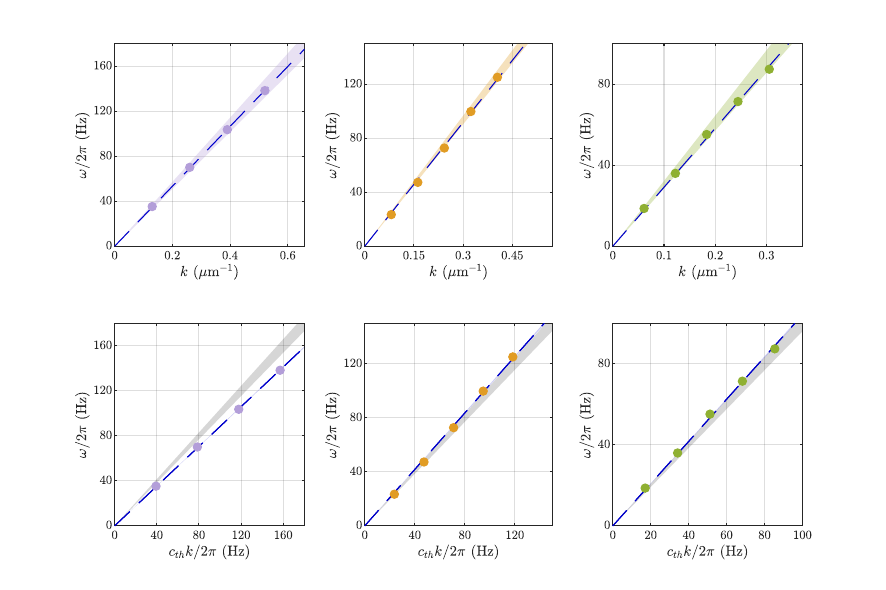}
\caption{\textbf{Dispersion relation from wavepackets measurements.} \textit{Upper row:} The excitation frequency $\omega/2\pi$ is plotted as a function of the momentum $k$. The dashed blue line represents a linear fit to the data, while the shaded region indicates NPSE simulations. Colors denotes different out-of-equilibrium protocols, consistent with Fig.~\ref{fig:main_damping_wp} of the main text: Low-amplitude wavepacket following a potential ramp (purple), edge-displacement (orange), and quench from a dimple to a flat box (green). The corresponding fitted speeds of sound are: $c_{r} = 1.674\,(0.005)\,\mathrm{\upmu m/ms}$, $c_{s}=1.923^{+0.008}_{-0.009}\,\mathrm{\upmu m/ms}$ and $c_{d} = 1.831^{+0.008}_{-0.009}\,\mathrm{\upmu m/ms}$, respectively, with error corresponding to the 68\% confidence interval, computed via bootstrap. 
\textit{Lower row:} The excitation frequency $\omega/2\pi$ is plotted against the expected value $\omega_{th}/2\pi=c_{th}k/2\pi$. The dashed blue line shows the linear fit, and the grey shaded region represents the expected range of $\omega_{th}$, reflecting the uncertainty in the measured linear density for each dataset. The expected speeds of sounds are respectively: $c_{th,r} = 1.89\,(0.07)\,$\si{\micro\meter\per\milli\second}, $c_{th,s} = 1.84\,(0.07)\,$\si{\micro\meter\per\milli\second}, $c_{th,d} = 1.76\,(0.07)\,$\si{\micro\meter\per\milli\second}.
Deviations between fitted and expected speeds of sound may arise from systematic uncertainties in the atom number, which is extracted from density ripples after 11.2 ms of ToF, whereas the longitudinal density profile is obtained from 2 ms of ToF images. Data are ordered and colour-coded as in the upper row. 
}
\label{fig_SM:dispersion_wp}
\end{figure*}

In Fig.~\ref{fig_SM:dispersion_shaking} and Fig.~\ref{fig_SM:dispersion_wp} we extract the dispersion relation by fitting the excitation frequencies $\omega_k$ as a function of the momenta $k$. We show the results for both single mode and wavepackets measurements. In all the cases, data are well described by a linear dependence, confirming that our experiments are conducted in the phononic regime. The corresponding speed of sound is obtained from the slope of the dispersion and compared to the theoretical expectation for a homogeneous system $c_{th} = \sqrt{g_n n_{1\mathrm{d}}/m}$, where $g_n$ is the effective density-dependent interaction strength~\cite{RauerThesis}.\\

\section{Preparation of the initial state.}
Our quasi-condensate of $^{87}$Rb atoms is prepared using standard protocols of magneto-optical trapping, laser cooling, and evaporative cooling to reach quantum degeneracy. The atom chip generates an rf-dressed, cigar-shaped magnetic trap, in which the final stage of evaporative cooling is performed \cite{RauerThesis}. The trapping frequency in the two tightly confined transverse directions is fixed, but has varied slightly over time due to setup upgrades and optimization. In the weak-perturbation experiments, $\omega_\perp = 2\pi \times 1.27$ kHz, whereas for medium- to strong-perturbation experiments, $\omega_\perp \simeq 2\pi \times 1.38$ kHz. By contrast, the longitudinal trapping frequency is the same for all datasets, $\omega_\parallel \simeq 2\pi \times 7$ Hz.
\\The geometry of the potential in the longitudinal direction can be modified by superimposing a blue-detuned dipole potential onto one of the two transverse directions of the magnetic trap. The box-like potential is realized using two different methods: either by projecting the blue-detuned laser onto an opaque mask \cite{Rauer2018, RauerThesis}, or by shaping it with a digital micromirror device (DMD) \cite{Tajik:19, Cataldini2022PRX}.
\\The total atom number in the initial state is controlled via the evaporative cooling radio frequency. The atom number is post-selected to achieve $\sigma_N / \bar{N} \approx 0.1$, where $\bar{N}$ denotes the mean atom number and $\sigma_N$ its standard deviation, thereby reducing shot-to-shot variations in the speed of sound arising from atom number fluctuations. Total atom number and temperature are extracted from density ripples observed after 11.2 ms of ToF. \cite{ManzPhysRevA.81.031610, SchweiglerThesis}.\\

\section{Box confinement using an opaque mask.}
The box trap employed in the single-mode excitation experiments (Fig.~\ref{fig:main_damping_shaking}) and small-amplitude wavepacket measurements (Fig.~\ref{fig:main_damping_wp}(b)) is realized by using a blue-detuned laser light of wavelength $\lambda_{\text{mask}}=767$ nm. To create the box, the laser light is shined onto a wired mask with opaque stripes of varying widths. The transversely aligned Gaussian beam, blocked at its centre by the selected stripe, generates a potential that adds two steep walls to the shallow harmonic magnetic confinement. In these experiments, the box length is $L \sim 49\,\upmu$m, with wall widths $w_{L/R} \simeq 2.5\,\upmu$m. Owing to the small box size relative to the longitudinal potential, the curvature of the bottom of the trap can be neglected.\\

\section{Box confinement using a DMD.}
The wave packet protocols realized using the DMD are shown in Fig.~\ref{fig:main_damping_wp}(c,d). In these experiments, the laser light employed to produce the box trap has a wavelength $\lambda_{\text{DMD}}=660$ nm. The DMD compensates for the harmonic curvature of the bottom of the trap~\cite{Tajik:19}, and provides the submicron accuracy, with a projected pixel size of $0.417\,\upmu$m in the plane of the atoms. Each trap geometry is obtained through an iterative optimization of the DMD mask against the measured density profile. Due to experimental limitations, the box walls acquire a finite width; fitting the density with error functions yields wall widths of $2$–$3\,\upmu$m. \\

\section{Out of equilibrium protocols.}
The dynamics of the system are probed through a statistically large set of absorption images after 2 ms of ToF, from which we extract the longitudinal density profile $n(z,t)$. Each data set usually includes 150 -- 250 repetitions. During the dynamics of our Bose gas, the heating of the system is negligible, and the measured atom loss rate ranges between 0.5 and 2 atoms/ms; it arises from three-body recombination, collisions with the background gas particles, and technical noise.\\

\begin{figure*}
\centering
\includegraphics[trim={0mm 1mm 0mm 0mm},clip, width=\linewidth]{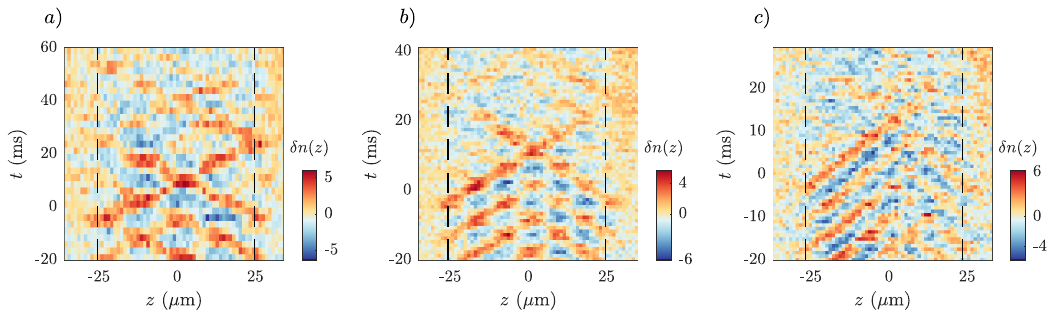}
\caption{\textbf{Excitation of higher order phononic modes.} The $4^{\mathrm{th}}$, $6^{\mathrm{th}}$ and $8^{\mathrm{th}}$ modes are addressed in panel (a), (b) and (c) respectively. Each 2D plot shows the evolution of the density perturbation profile $\delta n(z,t)$ during the last $20\,$ ms of shaking (negative times) and for the full subsequent evolution. By convention $t=0$ sets the end of the driving. The colorbar represents the perturbation amplitude in $\upmu$m$^{-1}$. The dashed black lines represent the position of the box walls which is used in the computation of the Fourier decomposition (see Fig.~\ref{fig:main_damping_shaking}(b-e))} 
\label{fig_SM:shaking_high_order_density_carpets}
\end{figure*}

\textit{a) Single mode excitation via resonant driving.}
The generic $j^{\mathrm{th}}$ phononic mode is excited by continuously modulating the intensity of the blue-detuned laser light at frequency $\omega_j=ck_j=c\pi j/L$, where $c$ is the speed of sound and $L$ is the box length. The modulation is generated through an arbitrary waveform generator as $V(-\tau_{dr}\le t\le 0) = V_0 + V_{\epsilon}\cos(\omega_jt)$, with $\tau_{dr}$ the driving time. This protocol is particularly suited for small perturbation amplitudes $V_\epsilon$, as stronger modulations cause non-negligible loss of atoms from the box, thus affecting $c$. The excitation of the lowest symmetric mode is shown in Fig.~\ref{fig:Density_carpets}(a), while the three next higher modes are shown in Fig.~\ref{fig_SM:shaking_high_order_density_carpets}. We consider relatively short shaking times, $\tau_{dr} = [30,\,60]\,$ ms. Driving the system for only a few oscillations limits unwanted heating and minimizes integrability-breaking processes such as three-body loss.\\

\textit{b) Single mode excitation via quench in the trapping potential.} In this protocol two potentials are independently designed and optimized with the DMD: a flat box potential and a box potential with an additional modulation at his bottom, $V_c\propto \cos(k_2z)=\cos(2\pi z/L)$, corresponding to the geometric shape of the second mode. The dynamics are initiated by rapidly quenching between these two configurations: The system is initially prepared in one potential and then quickly switched to the other. Importantly, within our parameter regime, the direction of the quench has no measurable influence on either the resulting dynamics or the damping rate~\cite{Cataldini2022PRX}. The quench is performed within a few microseconds, much faster than the characteristic timescale of the longitudinal dynamics.\\

\begin{figure}
\centering
\includegraphics[trim={0mm 0mm 0mm 0mm},clip]{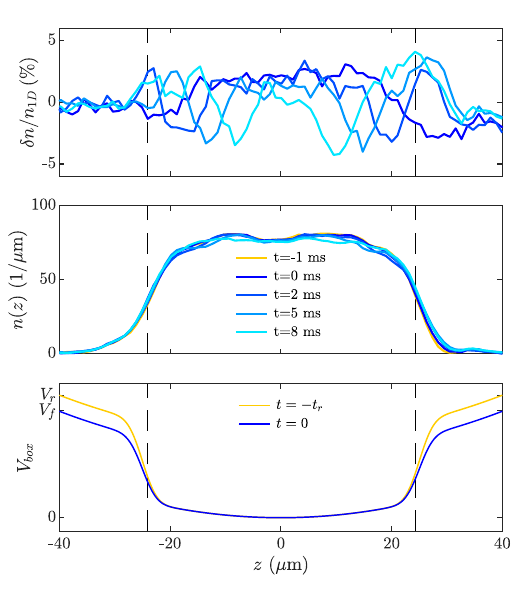}
\caption{\textbf{Ramp down of the potential.} Top: The density perturbation profile $\delta n(z)/n_{1\mathrm{d}}$ is plotted for selected times, $t=0, 2.5, 5, 7.5$ ms and shown with different gradients of blue, the darker corresponding to $t=0$ and the lightest corresponding to $t=7.5$ ms. Middle: The full longitudinal density profile $n(z)$ is shown for the same time steps as in the top panel. Additionally, the density profile for the thermal state is plotted with a yellow line. Bottom: Schematic of the experimental protocol, with $V_r$ representing the intial potential amplitude and $V_f = V_r-\delta V$ the final. In all the three panels, the black dashed lines indicate the position of the walls of the box.
Although the change in potential is barely noticeable in the full density profile as extracted from $2$ ms ToF pictures, the propagation of the perturbation becomes evident when the background profile $\langle n(z)\rangle_t$ is removed (upper panel). 
The initial thermal state has temperature $T_r=40\,(4)$ nK and linear density $n_{1\mathrm{d}}=78\,(8)\,\upmu$m$^{-1}$.
The complete evolution carpet is shown in Fig.~\ref{fig:Density_carpets}(b) of the main text, where the dynamics have been recorded for 100 ms, and the relative mode decomposition is plotted in Fig.~\ref{fig:main_damping_wp}(b).}
\label{fig_SM:wp_kick_scheme}
\end{figure}

\textit{c) Wave packets after fast ramp.}
A schematic of the protocol is illustrated in the panel at the bottom of Fig.~\ref{fig_SM:wp_kick_scheme}. The system, initially prepared in the potential $V_0$ is ramped down in $t_r = 1\,$ ms to the final value $V_f$. The small change in the position of the box walls creates a depletion of atoms which begins to propagate from the edges inward, at the speed of sound $c=1.7\,\upmu$m/ms (see Fig.~\ref{fig_SM:wp_kick_scheme} and Fig.~\ref{fig_SM:dispersion_wp}). The amplitude of the wave packets is about $5\%$ of the linear density.
This protocol  is used for the experiment shown in Fig.~\ref{fig:Density_carpets}(b), with the corresponding mode dynamics displayed in Fig.~\ref{fig:main_damping_wp}(b). \\

\begin{figure}
\centering
\includegraphics[trim={0mm 0mm 0mm 0mm},clip]{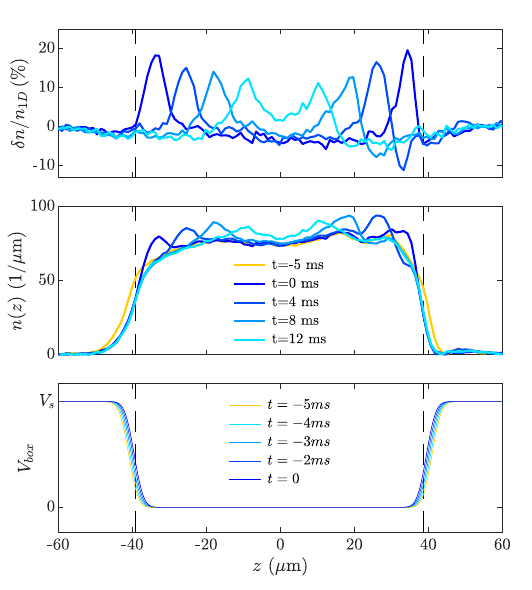}
\caption{\textbf{Edge displacement protocol.} Top: The density perturbation profile $\delta n(z)/n_{1\mathrm{d}}$ is plotted for selected times, $t=0, 4, 8, 12$ ms and shown with different shades of blue, the darker corresponding to $t=0$ and the lightest corresponding to $t=12$ ms. Middle: The full density profile $n(z)$ is shown for the same time steps as in the top panel. Additionally, the density profile for the thermal state is plotted with a yellow line. Bottom: Schematic of the experimental protocol: in yellow is shown the potential at thermal state ($t=-5\,$ms) and in dark blue the final potential ($t=0$). The intermediate potentials are shown in gradients. In all the three panels, the black dashed lines indicate the position of the walls of the box. The system is initially prepared in a thermal state at temperature $T = 60\,(6)$ nK and has a linear density of $n_{1\mathrm{d}}=78\,(8)\,\upmu$m$^{-1}$.}
\label{fig_SM:wp_shift_scheme}
\end{figure}

\textit{d) Box edge displacement.}
Starting from a box potential of length $L$, both walls are shifted in $n$ small steps of size $\Delta z$, and duration $\Delta t$, until it reaches the final box length $L_s$ (see Fig.~\ref{fig:main_damping_wp}(g)). This is achieved by sequentially turning on pixels rows of the DMD. During the translation of the box wall, a momentum is imprinted onto the atoms at the edge of the box and the propagation of a wave packet is initiated on each side, (Fig.\ref{fig_SM:wp_shift_scheme}). The bottom frame of Fig.\ref{fig_SM:wp_shift_scheme} schematically represents a measurement where the box length is reduced in $t_s = 5$ ms, by turning on one pixel row every millisecond (corresponding to $\Delta z\approx 0.4\,\upmu$m), yielding a total displacement of $\Delta L \approx 2\,\mathrm{\upmu}$m per wall, thus reducing the length of the box from $L = 82\,\upmu$m to $L_s = 78\,\upmu$m. The mechanism induces two counter-propagating wave packets, left (L) and right (R), with widths $w_L = w_R \sim 2 \upmu$m, amplitude $\delta n_L(t = 0) = \delta n_R(t = 0) \simeq 16 \,\upmu$m$^{-1}$, moving at a speed $v_L = v_R = 1.9\, \upmu$m/ms (see Fig.~\ref{fig_SM:wp_shift_scheme}). These correspond to a density perturbation of about $20\%$ of the linear density, $n_{1\mathrm{d}}=78\,\upmu$m$^{-1}$. This protocol is used for the experiment shown in Fig.~\ref{fig:wp_wave_breaking}(a), with the corresponding mode dynamics displayed in Fig.~\ref{fig:main_damping_wp}(c). \\

\begin{figure}
\centering
\includegraphics[trim={0mm 0mm 0mm 0mm},clip]{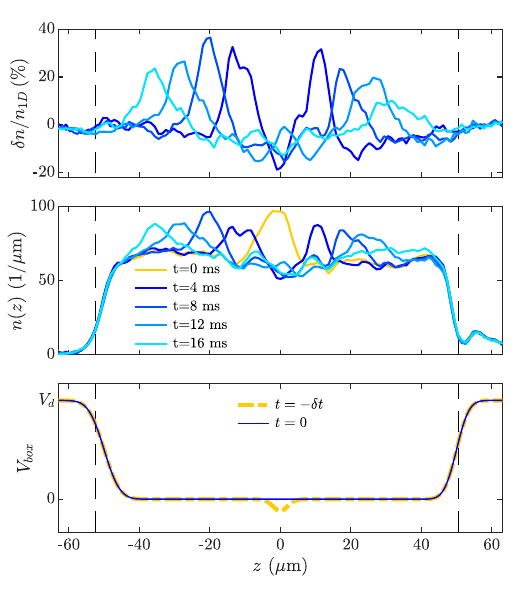}
\caption{\textbf{Dimple protocol.} Top: The density perturbation profile $\delta n(z)/n_{1\mathrm{d}}$ is plotted for selected times, the initial time $t=-\delta t$ is in yellow, while the evolution times $t=4, 8, 12, 16$ ms are shown with different shades of blue, the darker corresponding to $t=4$ ms and the lightest corresponding to $t=16$ ms. Middle: The full longitudinal density profile $n(z)$ is shown for the same time steps as in the top panel. Bottom: Schematic of the experimental protocol, where in yellow is shown the potential in the thermal state and in dark blue the flat potential it is quenched to. In all the three panels, the black dashed lines indicate the position of the walls of the box. In the initial state, the dimple has a width of $\sigma = 4\,\upmu$m, the temperature is $T=69\,(5)\,$ nK and the linear density $n_{1\mathrm{d}}=68\,(7)\,\mathrm{\upmu}$m$^{-1}$.}
\label{fig_SM:wp_dimple_scheme}
\end{figure}

\textit{e) Quench from dimple to flat box potential.}
A schematic of the protocol is illustrated in the panel at the bottom of Fig.~\ref{fig_SM:wp_dimple_scheme}. The initial state is prepared in a box potential of length $L_d=103\,\upmu$m, featuring a small dimple in the middle. The initial potential is suddenly quenched to a flat box potential, having same length $L_d$. Following the quench, the dimple splits into two wave packets, propagating from the middle to the edges of the box at speed $v_L \sim v_R \sim1.8\,\upmu$m/ms. 
The top and central panel of Fig.~\ref{fig_SM:wp_dimple_scheme} show the density perturbation $\delta n(z)$ and and the full density profile $n(z)$ respectively, during the first $16\,$ ms of evolution after the quench, in steps of $4\,$ ms. The thermal state at $t=-\delta t$ is plotted in yellow. Four milliseconds after the quench, their amplitude is $\delta n_L(t=4) = \delta n_R(t=4) = 27\,\upmu$m$^{-1}$, corresponding to $39\%$ of the linear density $n_{1\mathrm{d}}=68\,\,\mathrm{\upmu}$m$^{-1}$. The full dynamics of the perturbation are plotted in Fig.~\ref{fig:Density_carpets}(c) of the main text, with the corresponding mode dynamics displayed in Fig.~\ref{fig:main_damping_wp}(d).\\

\newpage
\clearpage

\section{Measurements list}
All the single mode measurements are listed in table~\ref{T_sm:shaking_meas_list} and~\ref{T_sm:PB_meas_list}, while all the wave packet measurements are listed in table~\ref{T_sm:wp_meas_list}.

The density carpets of Fig.~\ref{fig:Density_carpets} of the main text correspond to measurements 1S2, 3Wr and 12Wd respectively.
The mode decomposition carpet and resonant mode dynamics for measurements 1S2, 5S4, 7S6 and 9S8 is shown in Fig.~\ref{fig:main_damping_shaking} of the main text. Measurements from 1S2 to 9S8 are used to fit the damping in Fig.~\ref{fig:main_damping_shaking}(a). The crossover from weakly to highly nonlinear regime shown in Fig.~\ref{fig:gamma2_vs_perturbation_amp} includes measurements from 1S2 to 4S2 (red circles), 10S2, 11S2 (orange circles) and from 1Q to 15Q (blue diamonds). 
Measurements 3Wr, 6Ws and 12Wd are shown in Fig.~\ref{fig:main_damping_wp} of the main text. Fig.~\ref{fig:wp_wave_breaking} corresponds to measurement 6Ws.

\begin{table}[ht]
    \centering
    \begin{tabular}{|l|ccccccccc||cc|}
    \hline
    \rule{0pt}{3ex} \parbox{2cm}{\textbf{Code}} & \parbox{0.8cm}{\textbf{1S2}} & \parbox{0.8cm}{\textbf{2S2}} & \parbox{0.8cm}{\textbf{3S2}} & \parbox{0.8cm}{\textbf{4S2}} & \parbox{0.8cm}{\textbf{5S4}} & \parbox{0.8cm}{\textbf{6S4}} & \parbox{0.8cm}{\textbf{7S6}} & \parbox{0.8cm}{\textbf{8S8}} & \parbox{0.8cm}{\textbf{9S8}} & \parbox{1cm}{\textbf{10S2}} & \parbox{1cm}{\textbf{11S2}}\\ [0.4ex] 
    \hline \hline
    \rule{0pt}{3ex} $T\,$(nK)                           & 64  & 60  & 59  & 63  & 53  & 56  & 52  & 58  & 48  & 60  & 60  \\ [0.4ex] 
    \rule{0pt}{3ex} $\sigma_T\,$(nK)                    & 11  & 2   & 6   & 9   & 7   & 9   & 6   & 8   & 7   & 2   & 2   \\ [0.4ex]
    \rule{0pt}{3ex} $n_{1\mathrm{d}}\,$ $(\upmu$m$^{-1})$          & 73  & 57  & 70  & 62  & 77  & 69  & 73  & 60  & 74  & 57  & 57   \\ [0.4ex] 
    \rule{0pt}{3ex} $\sigma_{n_{1\mathrm{d}}}\,$ $(\upmu$m$^{-1})$ & 8   & 7   & 7   & 7  & 6   & 7   & 7   & 6   & 8   & 6   & 6   \\ [0.4ex]
    \rule{0pt}{3ex} $A_j\,$ $(\upmu$m$^{-1})$             & 3.8 & 2.4 & 5.4 & 2.3 & 3.5 & 2.7 & 3.0 & 1.9 & 2.7 & 5.1 & 6.9 \\ [0.4ex]
    \hline
    \end{tabular}
    \caption{\textbf{Single mode (resonant driving) measurements.} The measurements are listed with the code \textit{nSj}, where \textit{n} is a sequential number, \textit{S} stands for `shaking' and \textit{j} is the excited mode number. All the measurements in the table have box length $L\approx49\,\mathrm{\upmu m}$.}
    \label{T_sm:shaking_meas_list}
\end{table}

\begin{table}[h!]
    \centering
    \begin{tabular}{|l|ccccccccccccccc|}
    \hline
    \rule{0pt}{3ex} \parbox{2cm}{\textbf{Code}} & \parbox{0.8cm}{\textbf{1Q}} & \parbox{0.8cm}{\textbf{2Q}} & \parbox{0.8cm}{\textbf{3Q}} & \parbox{0.8cm}{\textbf{4Q}} & \parbox{0.8cm}{\textbf{5Q}} & \parbox{0.8cm}{\textbf{6Q}} & \parbox{0.8cm}{\textbf{7Q}} & \parbox{0.8cm}{\textbf{8Q}} & \parbox{0.8cm}{\textbf{9Q}} & \parbox{0.8cm}{\textbf{10Q}} & \parbox{0.8cm}{\textbf{11Q}} & \parbox{0.8cm}{\textbf{12Q}} & \parbox{0.8cm}{\textbf{13Q}} & \parbox{0.8cm}{\textbf{14Q}} & \parbox{0.8cm}{\textbf{15Q}} \\ [0.4ex] 
    \hline \hline
    \rule{0pt}{3ex} $T\,$(nK)                                      & 46  & 48  & 64  & 68 & 74  & 88 & 90 & 90 & 106 & 114 & 116 & 120 & 120 & 122 & 132 \\ [0.4ex] 
    \rule{0pt}{3ex} $\sigma_T\,$(nK)                               & 4   & 8   & 10  & 8  & 7   & 6  & 11 & 9  & 10  & 19  & 15  & 9   & 10  & 11  & 16  \\ [0.4ex]
    \rule{0pt}{3ex} $n_{1\mathrm{d}}\,$ $(\upmu$m$^{-1})$          & 60  & 73  & 60  & 69 & 83  & 80 & 76 & 80 & 91  & 85  & 90 & 61  & 78  & 71  & 71  \\ [0.4ex] 
    \rule{0pt}{3ex} $\sigma_{n_{1\mathrm{d}}}\,$ $(\upmu$m$^{-1})$ & 5   & 7   & 6   & 7  & 8   & 8  & 8  & 8  & 10  & 9   & 10  & 6   & 8   & 8   & 8   \\ [0.4ex]
    \rule{0pt}{3ex} $A_2\,$ $(\upmu$m$^{-1})$                      & 8   & 13  & 12  & 28 & 40  & 29 & 18 & 40 & 30  & 20  & 32  & 10  & 28  & 26  & 31  \\ [0.4ex]
    \hline
    \end{tabular}
    \caption{\textbf{Single mode (potential quench) measurements.} The measurements are listed with the code \textit{nQ}, where \textit{n} is a sequential number, \textit{Q} stands for `quench' from cosine potential to flat box. For all the measurements in the table the excited mode is the $j=2$ and they all have box length $L\approx80\,\mathrm{\upmu m}$.}
    \label{T_sm:PB_meas_list}
\end{table}

\begin{table}[h!]
    \centering
    \begin{tabular}{|l|c|c|c|c|c|c|c|c|c|c|c|c|}
    \hline
    \rule{0pt}{3ex} \parbox{2cm}{\textbf{Code}} & \parbox{0.8cm}{\cellcolor{ramp!30}\textbf{1Wr}} & \parbox{0.8cm}{\cellcolor{ramp!30}\textbf{2Wr}} & \parbox{0.8cm}{\cellcolor{ramp!30}\textbf{3Wr}} & \parbox{0.8cm}{\cellcolor{edgeshift!30}\textbf{4Ws}} & \parbox{0.8cm}{\cellcolor{edgeshift!30}\textbf{5Ws}} & \parbox{0.8cm}{\cellcolor{edgeshift!30}\textbf{6Ws}} & \parbox{0.8cm}{\cellcolor{dimple!30}\textbf{7Wd}} & \parbox{0.8cm}{\cellcolor{dimple!30}\textbf{8Wd}} & \parbox{0.8cm}{\cellcolor{dimple!30}\textbf{9Wd}} & \parbox{0.8cm}{\cellcolor{dimple!30}\textbf{10Wd}} & \parbox{0.8cm}{\cellcolor{dimple!30}\textbf{11Wd}} & \parbox{0.8cm}{\cellcolor{dimple!30}\textbf{12Wd}} \\ [0.4ex] 
    \hline \hline
    \rule{0pt}{3ex} $T\,$(nK)                                     &  47  & 53  & 40  & 54  & 70  & 60  & 64  & 64  & 69   & 64   & 64   & 69  \\ [0.4ex] 
    \rule{0pt}{3ex} $\sigma_T\,$(nK)                               & 6   & 5   & 4   & 6   & 8   & 6   & 7   & 7   & 5    & 7    & 7    & 5   \\ [0.4ex]
    \rule{0pt}{3ex} $n_{1\mathrm{d}}\,$ $(\upmu$m$^{-1})$          & 86  & 88  & 78  & 68  & 84  & 78  & 70  & 70  & 68   & 73   & 66   & 68  \\ [0.4ex] 
    \rule{0pt}{3ex} $\sigma_{n_{1\mathrm{d}}}\,$ $(\upmu$m$^{-1})$ & 8   & 8   & 8   & 8   & 9   & 8   & 7   & 7   & 7    & 7    & 6    & 7   \\ [0.4ex]
    \rule{0pt}{3ex} $L\,$($\mathrm{\upmu}$m)                       & 50  & 48  & 48  & 54  & 80  & 78  & 101 & 101 & 103  & 97   & 101  & 103 \\ [0.4ex]
    \hline
    \end{tabular}
    \caption{\textbf{Wave packets measurements.} The measurements are listed with the code \textit{nWx}, where \textit{n} is a sequential number, \textit{W} stands for `wave packet' and \textit{x} refers to the protocol type \textit{x=r} for `ramp', \textit{x=s} for `edge shift' and \textit{x=d} for `dimple quench'. The color scheme is used for clarity and is consistent with Fig.~\ref{fig:main_damping_wp} of the main text.}
    \label{T_sm:wp_meas_list}
\end{table}

\section{Power-law modeling of the damping rate}
The behaviour of the normalized damping rate $\tilde{\Gamma}(\tilde{k})$, with $\tilde{\Gamma} = \frac{m\lambda_T^2}{4\hbar}\Gamma$ and $\tilde{k}=\lambda_T k/2$, extracted from single mode excitation (Fig.~\ref{fig:main_damping_shaking} of the main text) is modeled using a power-law function of the form $y=\alpha x^\beta$. To linearize this model, both variables are log-transformed, giving $\log(y) = \log(\alpha)+\beta\log(x)$. A linear regression is then performed in log space, with centering applied to the predictor variable to improve numerical stability and to reduce correlations. The fitted slope $\beta$ represents the scaling exponent, while the intercept corresponds to $\log(\alpha)$, allowing estimation of the original parameters in the power-law relationship. The estimated parameters in the original $(\tilde{\Gamma},\tilde{k})$-space are $\beta = 1.48\,(0.03)$ and $\alpha=0.76\,(0.02)$, with uncertainties that give the standard deviations derived from the covariance matrix of the fit.

\begin{figure}[h!]
    \centering
    \includegraphics{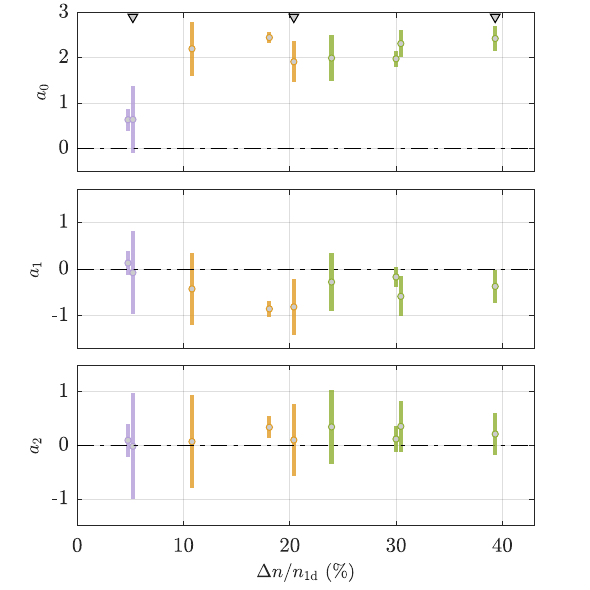}
    \caption{\textbf{Polynomial regression on wave packets measurements.}
    The ratio $\Theta=\Gamma/\Gamma_{th}$, with $\Gamma$ and $\Gamma_{th}$ the measured and predicted damping rate, is fitted with up to second-order Legendre polynomials $f=a_0P_0+a_1P_1+a_2P_2$. The statistical significance of the coefficients $a_0$, $a_1$ and $a_2$ is evaluated via polynomial regression. The extracted coefficients are plotted as a function of the relative perturbation strength $\Delta n/n_{1\mathrm{d}}$, as defined in the main text. The errors on the coefficients are obtained from the fit. In purple: data sets corresponding to ramp/up of the box potential; in orange: edge displacement measurements; in green: dimple quenched to a flat box potential. The three grey arrows on top mark the data sets shown in the main text.
    The measurements have $n_{1\mathrm{d}}\in[65-90]\,\upmu$m$^{-1}$, $T\in[40,70]\,$nK.}
    \label{fig_SM:poly_regression}
\end{figure}

\section{Polynomial regression analysis}
To assess the agreement between the damping rate measured via wave packet excitations $\Gamma$, and the theoretical prediction $\Gamma_{th}=\sqrt{(k_BT)/(mn_{1\mathrm{d}})}\,k^{3/2}$, we perform polynomial regression analysis on the ratio $\Theta(k) = \Gamma(k)/\Gamma_{th}(k)$. We fit $\Theta$ with $f=a_0P_0(\xi) + a_1P_1(\xi)+a_2P_2(\xi)$, where $P_0,P_1,P_2$ are the zeroth, first and second order Legendre polynomials respectively, defined with respect to the center of the $k$-axis, $\xi = (k-\bar{k})/\Delta k$, with $\bar{k} = (k_{min}+k_{max})/2$ and $\Delta k=k_{max}-\bar{k}$, and evaluate the statistical significance of the coefficients $a_0,\,a_1,\,a_2$. The zeroth-order coefficient $a_0$ represents a uniform scaling factor between experimental data and theory, while the higher-order coefficients, $a_1\,a_2$, capture systematic deviations as a function of $\xi$. In addition to the three data sets analyzed in Fig.~\ref{fig:main_damping_wp} of the main text, we perform a polynomial regression analysis on additional measurements where we excite wave packets with varying perturbation amplitude, $\delta n/n_{1\mathrm{d}}\in[0.05,0.4]$. The results are plotted in Fig.~\ref{fig_SM:poly_regression}. Here, we show the fitted coefficients $a_0,\,a_1,\,a_2$ with relative errorbars. Data points with distinct colors correspond to different kinds of protocol: purple for ramp up/down of the potential, orange for box edge displacement, and green for dimple quenched to a flat box. The data presented in Fig.~\ref{fig:main_damping_wp} of the main text, are here highlighted by grey arrows at the top of the figure. For small perturbation measurements, $a_1,\,a_2\approx0$ and only the zeroth-order coefficient $a_0$ is statistically significant, consistent with $\Theta$ being constant, and that the experimental data agree with the theoretical model. Additionally, the scaling factor close to $0.7$, is consistent with the coefficient $\alpha=0.76$ obtained by fitting the shaking measurements damping rate (Fig.~\ref{fig:main_damping_shaking} of the main text). For perturbation amplitude $\delta n/n_{1\mathrm{d}} \geq 0.16$, the higher-order coefficients $a_1,\,a_2$ are statistically significant, indicating a deviation from Andreev's prediction. \\
Next, to further corroborate the polynomial regression results within errorbars, we perform a $\chi^2$ hypothesis test: we assume to be true the null hypothesis $\frac{\Gamma} {\Gamma_{th}} = a_0$ and compute, for each data set the $\chi^2 = \sum_{i=1}^N (\Gamma_i-a_0P_0)^2/\sigma_{\Gamma_i}^2$, with $N$ number of excited modes per data set and $\sigma_{\Gamma_i}$ errors on damping, indicating the $68\%$ of confidence interval, computed via bootstrap. The results are summarized in the table below~\ref{T_sm:hypotesis_test}, where $\chi_0^2$ denotes the critical $\chi^2$ value corresponding to a $5\%$ significance level. Of all the data sets considered, only those corresponding to the lowest perturbation amplitudes ($5\%$ and $7\%$), satisfy $\chi^2<\chi_0^2$, indicating that the null hypothesis cannot be rejected in these cases and therefore, that they follow Andreev's theoretical prediction within their experimental uncertainties.
\begin{table}[h!]
    \centering
    \begin{tabular}{|l|c|c|c|c|c|c|c|c|c|}
        \hline
        \rule{0pt}{3ex} \parbox{2cm}{\textbf{Code}}   & \parbox{1cm}{\cellcolor{ramp!30}\textbf{1*Wr}} & \parbox{1cm}{\cellcolor{ramp!30}\textbf{3Wr}} & \parbox{1cm}{\cellcolor{edgeshift!30}\textbf{4Ws}} & \parbox{1cm}{\cellcolor{edgeshift!30}\textbf{5Ws}} & \parbox{1cm}{\cellcolor{edgeshift!30}\textbf{6Ws}} & \parbox{1cm}{\cellcolor{dimple!30}\textbf{7*Wd}} & \parbox{1cm}{\cellcolor{dimple!30}\textbf{9Wd}} & \parbox{1cm}{\cellcolor{dimple!30}\textbf{10*Wd}} & \parbox{1cm}{\cellcolor{dimple!30}\textbf{12Wd}} \\ [0.4ex]  
        \hline \hline
        \rule{0pt}{3ex} $\delta n/n_{1\mathrm{d}}\,(\%)$ & 5   & 5 & 11 & 18  & 20   & 24 & 30 & 31 & 39 \\ [0.4ex] 
        \rule{0pt}{3ex} $\chi^2$                & 10  & 5 & 45 & 126 & 332  & 89 & 12 & 48 & 56 \\ [0.4ex] 
        \rule{0pt}{3ex} $\chi_0^2$              & 13  & 8 & 9  & 11  & 9    & 14 & 9  & 16 & 9  \\ [0.4ex] 
        \hline
    \end{tabular}
    \caption{\textbf{$\chi^2$ Hypothesis Test under the Null Hypothesis $\Gamma/\Gamma_{th} = a_0$.} For each perturbation amplitude, the computed $\chi^2$ value is listed alongside the critical value $\chi_0^2$, at a significance level of $5\%$. The null hypothesis cannot be rejected when $\chi^2<\chi_0^2$. To improve the quality of the analysis, data sets with very close temperature and perturbation amplitude have been grouped: measurements 1Wr and 2Wr are grouped into 1*Wr, 7Wd and 8Wd are grouped into 7*Wd and measurements 10Wd and 11Wd are grouped into 10*Wd (see table~\ref{T_sm:wp_meas_list}). The color scheme is used for clarity and is consistent with Fig.~\ref{fig_SM:poly_regression}.}
    \label{T_sm:hypotesis_test}
\end{table}

\section{Theoretical background}
An ultracold 1D gas made of bosons with mass $m$, chemical potential $\upmu$, trapped in a longitudinal potential $U(z)$ is described, in mean-field approximation, by the Hamiltonian:
\begin{equation}
    H = \int dz\,\psi^\dagger\bigg[-\frac{\hbar^2}{2m}\partial_z^2  + U(z)-\upmu + \frac{g_{1\mathrm{d}}}{2} \psi^\dagger \psi \bigg]\psi,
    \label{SEQ:1d_ham}
\end{equation}
where $z$ is the longitudinal direction of the gas, $g_{1\mathrm{d}}$ is the 1D coupling constant characterizing the interparticle interaction, here assumed to occur via a $\delta$-function. For repulsive interactions $g_{1\mathrm{d}} > 0$. The Hamiltonian (\ref{SEQ:1d_ham}) without the term containing $U(z)$ and $\upmu$ is the Lieb-Liniger Hamiltonian \cite{LL1, LL2}. For brevity, the dependence on $z$ has been removed, $\psi^\dagger(z)\equiv \psi^\dagger$, $\psi(z)\equiv\psi$. The Bogoliubov theory can be extended to 1D systems \cite{MoraCastin}, in phase-density representation, assuming the wavefunction to be expressed as $\psi(z,t) = \sqrt{\rho(z,t)}\,e^{i\theta(z,t)}$, where $\rho$ and $\theta$ represent the local density and the phase of the quasicondensate respectively. After expanding the density around a uniform background $\rho(z,t) = \rho_0 +\delta\rho(z,t)$ and assuming small density fluctuations $\delta\rho(z,t) \ll |\rho_0|$, $H$ can be expanded in terms of the phase $\theta$ and the density fluctuations $\delta \rho$: $H = H_1 + H_2 + H_3 + \dots $. The Hamiltonian $H$ can be expressed as:
\begin{equation}
    H \approx \int dz\,\bigg[\frac{\hbar^2\rho_0}{2m}(\partial_z\theta)^2 + \frac{g_{1\mathrm{d}}}{2}(\delta\rho)^2 + \frac{\hbar^2}{8m\rho_0}(\partial_z\delta\rho)^2
    +\frac{\hbar^2}{2m}\delta\rho(\partial_z\theta)^2-\frac{\hbar^2}{8m\rho_0^2}\delta\rho(\partial_z\delta\rho)^2\bigg].
    \label{SEQ:LL_expansion}
\end{equation}
In the former equation, the first two quadratic terms represent the Luttinger-liquid Hamiltonian, which does not capture the full long-time scale dynamics of a weakly-interacting 1D Bose gas out of equilibrium \cite{Cataldini2022PRX,Rauer2018}. The third term in eq.~(\ref{SEQ:LL_expansion}) is the quantum pressure, and the last two terms represent, respectively, the phonon-phonon interaction and the density fluctuations self-interaction. In the phononic regime, dominated by long-wavelength excitations, the Bogoliubov dispersion relation becomes linear, and the leading interaction is phonon merging/branching. For $\omega = c|k|$ and at low $k$, Andreev has derived the power-law decay for the damping of phonons $\Gamma\propto |k|^{3/2}$ \cite{Andreev1980}. In this regime the quantum pressure term is also negligible, while it is responsible for the wave-breaking mechanisms occurring for larger perturbations. 

\section{Comparison to nonlinear Luttinger liquids}

The seminal book by Giamarchi \cite{Giamarchi2003} provides the foundational framework of one-dimensional quantum systems through bosonization and Luttinger liquid theory, where excitations are strictly stable due to the assumption of a linearized dispersion. Within this paradigm, correlation functions display universal algebraic behavior but phonons remain infinitely long-lived. The role of band curvature and other irrelevant operators is acknowledged, yet only as perturbative corrections within the renormalization group language, without a quantitative description of phonon lifetimes. Subsequent developments went beyond this framework: Samokhin \cite{Samokhin1998} showed that dispersion curvature leads to finite lifetimes of excitations, while the nonlinear Luttinger liquid (NLL) theory of Imambekov and Glazman \cite{Imambekov2009} established a universal description of dynamical response functions in the presence of nonlinearity, predicting the replacement of $\delta$-peaks by power-law edge singularities. 

Whereas Andreev’s self-consistent hydrodynamic approach predicts a finite phonon lifetime with the non-analytic scaling law $\Gamma_k \propto T^{1/2}k^{3/2}$ \cite{Andreev1980}, the nonlinear Luttinger liquid (NLL) framework developed by Imambekov and Glazman \cite{Imambekov2009} arrives at a qualitatively different picture. In the linear Luttinger model, phonons are strictly stable and appear as $\delta$-function peaks in dynamical response functions. The inclusion of dispersion curvature, however, removes this protection: in the NLL theory the sharp $\delta$-peaks are replaced by universal power-law edge singularities at the exact Lieb–Liniger excitation thresholds. Thus, rather than acquiring a simple exponential damping rate, phonons in the NLL picture manifest as broadened, asymmetric continua in $S(k,\omega)$ and $A(k,\omega)$. Our measurements directly probe the low-$k$, finite-temperature hydrodynamic regime where Andreev’s law is valid, while extensions toward higher-momentum or higher-energy excitations would connect to the universal non-Lorentzian lineshapes predicted by the NLL theory.

\end{bibunit}

\end{document}